\documentstyle[12pt,frascatiphys,epsfig]{article}
\def\cexe    { \pi^- p \rightarrow \eta \pi^+ \pi^- n }
\def\cexo    { \pi^- p \rightarrow \omega \omega n }
\def\gev     { GeV }

\def\tprime  { t^{\prime} }

\def\jpc     { J^{PC}}
\def\epp     { \eta \pi^+ \pi^- }
\def\ret     { \rho(770) \eta }
\def\atpi    { a_2(1320) \pi  }
\def\fte     { f_2(1270) \eta  }
\def\omom    { \omega\omega }
\begin{document}
\title{NATURAL PARITY RESONANCES 
 IN THE $\epp$ AND $\omom$
}
\author{
Valeri Dorofeev, VES collaboration \\
{\em IHEP, Protvino, Russia, RU-142284} \\
David Bugg, \\
{\em QMWC, Mile End Rd., London E1 4NS, UK}
}
\maketitle
\baselineskip=14.5pt
\begin{abstract}
Results of the $\epp$ and $\omom$ final state partial wave analysis(PWA) 
are presented.
The mass-independent and the mass-dependent 
PWA were performed.
The $\epp$ system PWA results in the observation of the resonances in the
$\jpc = 1^{--}$, $3^{--}$, $2^{++}$ and $4^{++}$ waves. 
The results of the $\omom$ system PWA confirms the existence of
the $f_2(1565)$ and $f_4(2050)$. The $f_2(1950)$ is observed.
There is evidence for a $f_4(2300)$ meson.
\end{abstract}
\baselineskip=17pt
%

\section {Introduction}
The $\epp$ and $\omom$ final states has been studied in the 
charge-exchange  reactions
\begin{equation}
\cexe , \quad \eta \rightarrow \gamma \gamma
\label{reac}
\end{equation}
and
\begin{equation}
\cexo , \quad \omega \rightarrow \pi^+\pi^-\pi^0.
\label{reaco}
\end{equation}
respectively.
Our previous results of the analysis of reaction (\ref{reac})
were reported at the conference \cite{dima97} 
and of reaction (\ref{reaco}) were published in \cite{ves2}.
The measurements were carried out using VES spectrometer \cite{bec1} exposed
by the $37 \, \gev$ momentum $\pi^-$ beam.

The study of the final system is done in two steps.
At first, the mass-independent fit 
of events, sorted into the narrow mass bins is carried out. 
Within a bin, the Dalitz plot and
the angular distributions are fitted to individual partial waves
of all possible coupling of the isobar spins and their orbital angular momentum.
In the second stage the mass-dependent fit is done inside the wide
mass band. Each partial wave production amplitude is saturated by the coherent 
sum of the resonance Breit-Wigner amplitudes.
\section {Results of the $\epp$ system analysis}
\begin{figure}[htb]
\epsfig{file=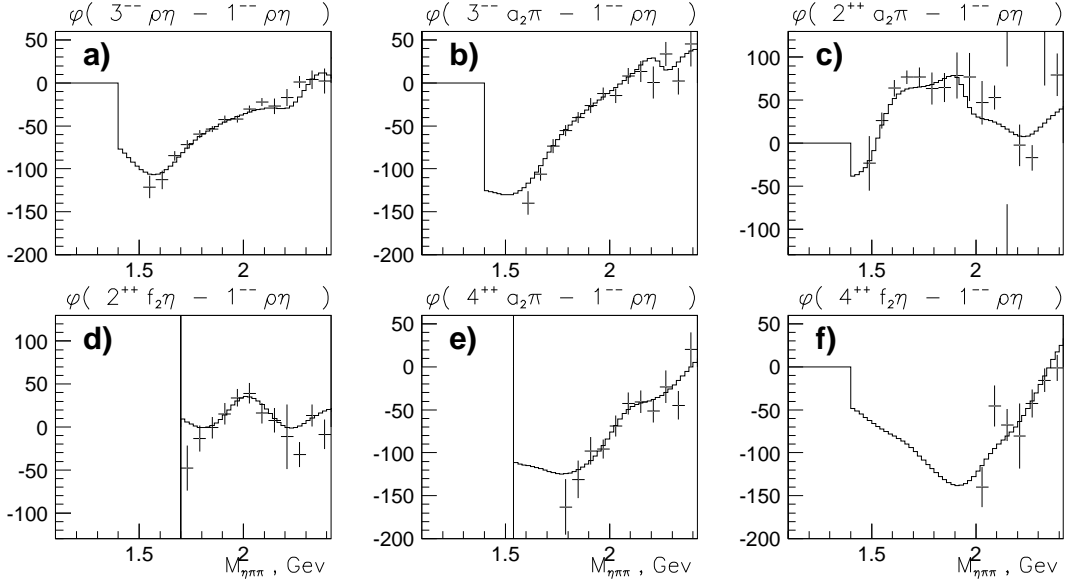,width=\textwidth,%
bbllx=0pt,bblly=0pt,bburx=565pt,bbury=320pt}
\caption{ The phases of the dominant waves.
The crosses are the results of the mass-independent PWA,
the curves - the mass-dependent PWA.   }
\label{phases1}
\end{figure}

We present the results of a
spin-parity analysis  of $4.2 \cdot 10^4  $ events of the 
$\epp$ final state in the mass range $1.4 \div 2.48 \, \gev$  for
$|\tprime| < 0.2 \, \gev^2$
\footnote{ $\tprime = t - t_{min}$, where $t$- momentum transfer from the beam to the
final state squared, $t_{min}$-its minimum value.}.
The $\epp$ system in this kinematical region is
dominated by the states with quantum numbers
allowed for the $ \pi \pi $ system:
 $J^P = 1^-$, $3^-$ with $I = 1$,
 and $J^P = 2^+$, $4^+$ with $I = 0$, due to the dominance of one pion exchange.
The model with the rank two density matrix \cite{chtrue} is used in the fit.
The results of the amplitude analysis are presented in
Fig.~\ref{intens1} and \ref{phases1}.
The $J^{PC}M^{\eta}=1^{++}0^+$ 
\footnote{notations from \cite{hansen}}
 $\ret$ S-wave is dominated in the natural 
parity exchange(NPE) set of the partial waves suppressed by the $\tprime$-cut.
In the unnatural parity exchange(UPE) sector the most significant waves are as follows:

\begin{figure}[htb]
\epsfxsize=15cm
\epsfysize=12cm
\epsfbox{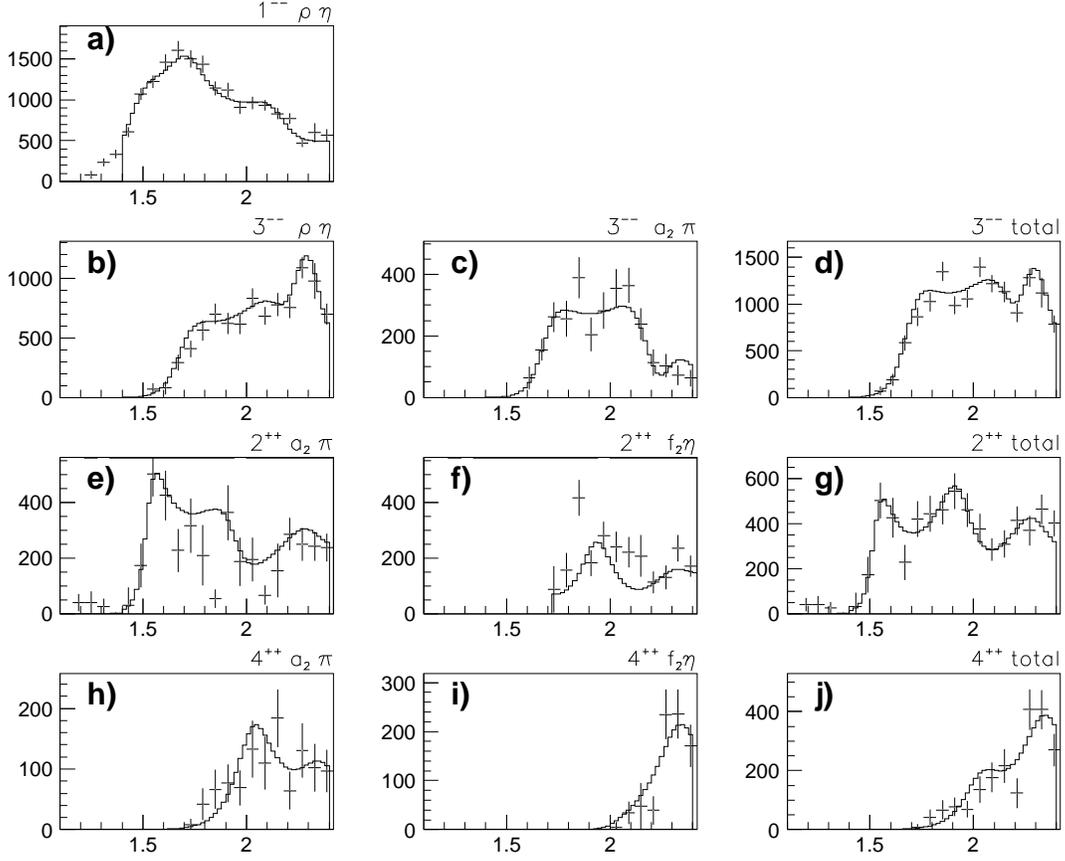}
\caption{ The  intensities for the UPE waves as a function of
$\epp $ mass in $\gev$. In crosses are shown the results of the mass-independent PWA,
in curves - of the mass-dependent PWA.   }
\label{intens1}
\end{figure}
\begin{figure}[htb]
\epsfig{file=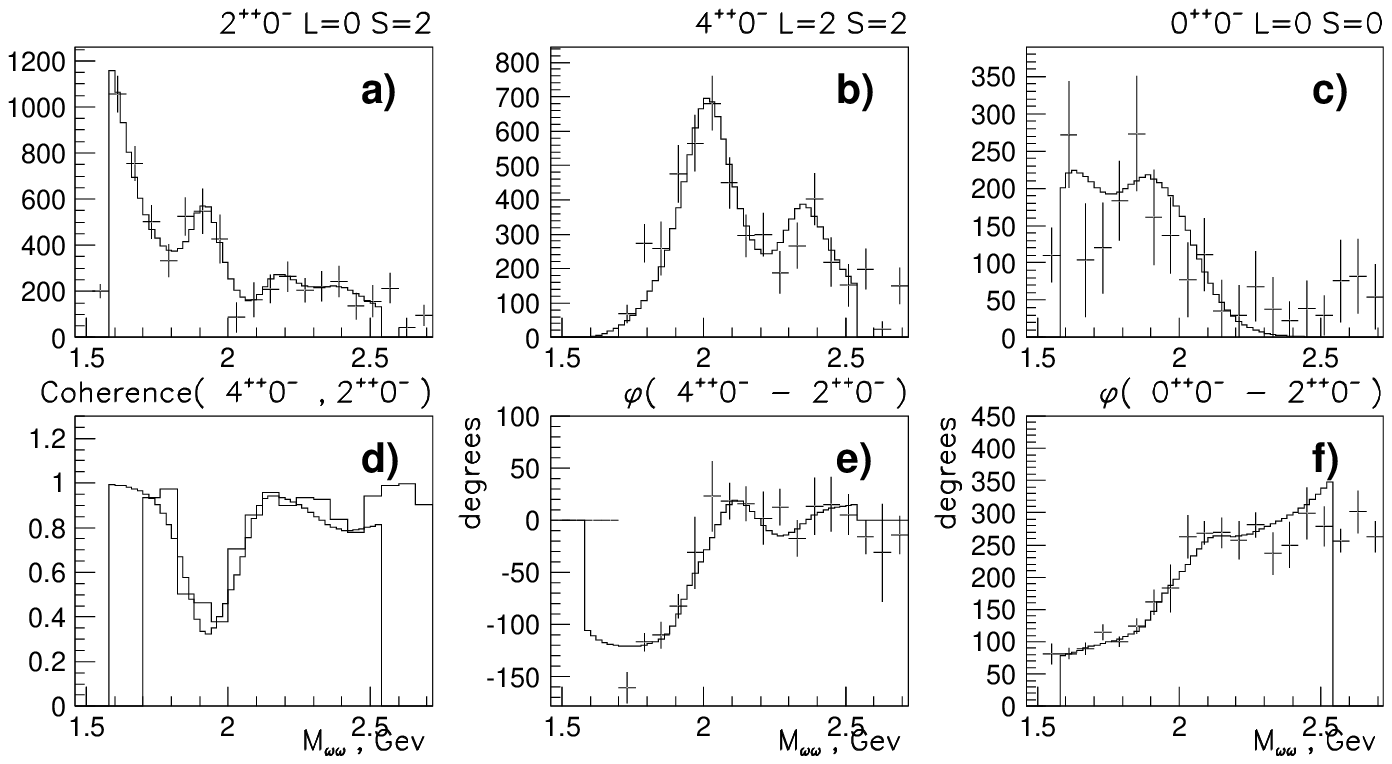,width=\textwidth}
\caption{a)-c) Intensities of the partial waves
(crosses - results of the mass-independent fit, 
curves - results of the mass-dependent fit). 
d) Coherence parameter between $4^+$ and $2^+$ waves. 
e) relative phase between $4^+$ and $2^+$ waves.
f) relative phase between $0^+$ and $2^+$ waves.}
\label{figww2}
\end{figure}
$\jpc = 1^{--}$.
The $1^{--} \ret $ wave is the dominant wave in the UPE sector, 
see Fig.~\ref{intens1}(a).
The best description is obtained when the $ 1^{--} $ wave is 
introduced as a coherent sum of the Breit-Wigner amplitudes
with PDG \cite{PDG} parameters for $ \rho(1450) $ , $ \rho(1700) $
and $ \rho(2150) $.
Amongst these, the $\rho (1700)$ is strongly dominant.
However, details of the $1^{--}$ wave are somewhat sensitive to
(i) the magnitude of the $\rho (1450)$ contribution and its poorly known
mass and width, (ii) the mass and width of the $\rho (2150)$.
This prevents us from
assigning accurate model independent parameters to  $\rho (1700)$.

$\jpc = 3^{--}$.
A clear result is the presence of the $ \rho_3(1690) $ decaying to the
$\atpi $ and $ \ret$; it is seen as a  rapid rise of the $3^{--}$
intensities in the $1.7 \, \gev$ region for both $\atpi $ and
$ \ret$, Figs.~\ref{intens1}(b) and (c),  and even more distinctly in
their phase variation relative to the $ 1^{--} $ wave, Figs.~\ref{phases1}(a) and (b).
The mass-dependent PWA gives  the following values
for the $ \rho_3(1690) $ mass and width:
$$
\begin{array}{ll}
M=1.69\pm 0.01(stat)~\pm 0.02(syst) \, \gev, & 
\Gamma=0.23\pm 0.02(stat)~\pm 0.05(syst) \, \gev.
\end{array}
$$
 Taking into account the PDG value of the branching ratio of the
$ a_2(1320) \rightarrow \eta \pi $ decay, we obtain:
$
 \frac{ BR (\rho_3(1690) \rightarrow \atpi)}
      { BR (\rho_3(1690) \rightarrow \ret )}
   = 6.0 \pm 3.0.
$
The error is mostly systematic.
There is an indication of a further resonance in 
the $ 3^{--} $  waves around $2.1 \, \gev$.
The $\atpi$ cross section falls sharply at about $2.15 \, \gev$ and
is fitted with a resonance at $M = 2.18 \pm 0.04 \, \gev$ with 
$\Gamma = 0.26 \pm 0.05 \, \gev$.
There is a corresponding weak peak in the $3^{--}$ total cross section 
summed over all final states, Fig.~\ref{intens1}(d).
The large phase variation observed in Figs.~\ref{phases1}(a) and (b) with respect to
the $1^-$ waves in the mass range 1.8 to $2.4 \, \gev$
requires that at least one $3^-$ resonance above $\rho _3(1690)$ 
is present in this mass range. 
At higher mass, there is some tentative evidence for further activity
around $2.3 \, \gev$, namely a peak in the $\atpi$ wave intensity and 
$3^{--}$ total cross section. A fit results in the following parameters of the
$\rho_3(2300)$:
$$
\begin{array}{ll}
M=2.30\pm 0.05 \, \gev, &  \Gamma=0.24 \pm 0.06\, \gev.   
\end{array}
$$

$\jpc = 2^{++}$.
The intensities of $ 2^{++} $ waves are rather small compared
to the $ 1^{--} $ and $ 3^{--} $, see Figs.~\ref{intens1}(e)-(g).
A sharp structure near $1.55 \, \gev$ is seen in the $ \atpi $
wave, Fig.~\ref{intens1}(e). The Breit-Wigner parameters found for this peak
are:
$$
\begin{array}{ll}
M=1.53\pm 0.02(stat) \pm 0.02(syst) \, \gev, &
\Gamma= 0.13\pm 0.03(stat) \pm 0.04(syst) \, \gev.
\end{array}
$$
We interpret it as an evidence for
the $f_2(1565)$ decaying through a previously unobserved $\atpi$ decay mode.
Considering now higher masses, the $ 2^{++} \fte $ intensity shows a
definite peak centered near $1.9 \, \gev$,  Fig.~\ref{intens1}(f).
Earlier, both VES \cite{ves2} and GAMS collaborations \cite{gams1}
have reported a $2^+$ resonance at $1.92 \, \gev$ in the $\omom$.
The $f_2(1950)$ parameters are fixed to that from the $\omom$-system fit:
$M = 1.93 \, \gev$, $\Gamma = 0.21 \, \gev$. 
A definite phase variation consistent with this mass and width is observed 
in Fig.~\ref{phases1}(d).                   

$\jpc = 4^{++}$.
The $\fte $
intensity shows a rapid rise in the $2.3 \, \gev$ region ( Fig.~\ref{intens1}(i) )
and have significant
phase variation in the region $2.0 \div 2.3 \, \gev$, Figs.~\ref{phases1}(f).
This may point to the existence of the $\fte $ resonance state.
If the parameters of the $f_4(2050) $ are kept at the PDG values, 
the following parameters for the $f_4(2300) $  meson are obtained:
$$
\begin{array}{ll}
M=2.33\pm 0.03 \, \gev, &  \Gamma=0.29 \pm 0.07\, \gev.   
\end{array}
$$

\section{Results of the $\omom$ analysis}
The results of the amplitude analysis of 9800 events of the $\cexo$
with $\omom$ mass inside $1.6 \div 2.5 \, \gev$ are shown in  Fig.~\ref{figww2}.
Production of the $0^-$, $1^+$, $2^-$ and $3^+$ final states requires weak NPE, 
amounting to $\sim 10\%$ of UPE.
The $2^{++}$ and the $4^{++}$ UPE waves are the most significant.
The production of the $0^+$ and $4^+$ final states turns out to be coherent,
corresponding to the $\pi$-exchange and spin-flip of the spectator nucleon. 
In the last stage of analysis these waves are included coherently.
The $2^{++}$ S-wave exhibits the violation of coherence with
the $4^{++}$ wave in the region of $1.9 \, \gev$, as shown
by the coherence parameter \cite{hansen} in Fig.~\ref{figww2}(d).
The reason for this is not understood now, although the
NPE $a_1$ exchange may be the possible explanation.
For this reason the model with the rank two density matrix 
is used in the fit, with the $2^{++}$ S-wave is presented in the second rank. 
There is a strong $2^+$ amplitude near the $\omom$ threshold, which 
may be attributed to the $f_2(1565)$ \cite{PDG}.
The mass region $1.9 \div 2.05 \, \gev$ requires two resonances: 
an $f_2$ with $M = 1.94 
\pm 0.01 \, \gev$, $\Gamma = 0.15 \pm 0.02 \, \gev$, and an $f_4$ with 
$M = 2.01 \pm 0.02 \, \gev$, $\Gamma = 0.25 \pm 0.04 \, \gev$.
In the higher mass range there is weak evidence for the
existence of a $f_4(2300)$ with the following parameters:
$$
\begin{array}{ll}
M=2.33\pm 0.02 \, \gev, &  \Gamma=0.24 \pm 0.04\, \gev.   
\end{array}
$$
\section{Conclusions}
Results of the $\epp$ and $\omom$ final state PWA 
reveal the resonant structures, which
are identified by the phase motion as well as peaks in the intensities of 
the partial waves. 
A fit to $1^{--} \ret$ partial wave is consistent with
the presence of a small contribution from the $\rho (1450)$, 
a dominant contribution from the $\rho (1700)$ with PDG values of the mass 
and width and some contribution from the $\rho (2150)$.
The well known $\jpc=3^{--} \rho _3(1690)$ is accompanied by a
$\rho_3(2300)$ in the $\epp$ final state. 
The  $f_2(1565)$ meson is observed in the $ \epp $ and $\omom$ final states. 
The $f_4(2050)$ is observed in the $\omom$ state.
There is indication of a $\rho_3(2100)$ in the $\epp$ and the $f_2(1950)$ in 
the $\epp$ and $\omom$. 
There is evidence for a $f_4(2300)$ meson, decaying to the 
$\fte$ in the F-wave and to the $\omom$.
%


\end{document}